\documentclass[superscriptaddress,showpacs,twocolumn,aps,epsfig]{revtex4}

\usepackage{graphicx}
\usepackage{dcolumn}
\usepackage{bm}
\usepackage[normalem]{ulem}
\usepackage{color}
\usepackage{amssymb}
\usepackage{amstext}

\begin{document}

\title{ Dipole-active optical phonons in YTiO$_3$: ellipsometry study and lattice-dynamics calculations}

\author{N. N. Kovaleva}
\author{A. V. Boris}
\affiliation{Max-Planck-Institut f\"{u}r Festk\"{o}rperforschung, Heisenbergstrasse 1, D-70569 Stuttgart, Germany}
\affiliation{Department of Physics, Loughborough University, Loughborough,
LE11 3TU, UK}
\author{L. Capogna}
\affiliation{CNR-INFM, CRS-SOFT and OGG Grenoble, 6 Rue J. Horowitz, BP
156 F-38042 Grenoble CEDEX 9, France}
\affiliation{Institut Laue Langevin, 6 rue J. Horowitz BP 156 38042 Cedex
9 Grenoble, France}
\author{J. L. Gavartin}
\affiliation{Accelrys, Cambridge, CB4 0WN, UK}
\author{\mbox{P. Popovich}}
\author{P. Yordanov}
\affiliation{Max-Planck-Institut f\"{u}r Festk\"{o}rperforschung, Heisenbergstrasse 1, D-70569 Stuttgart, Germany}
\author{A. Maljuk}
\affiliation{Hahn-Meitner-Institute, Glienicker Str. 100, D-14109 Berlin,
Germany}
\author{A. M. Stoneham}
\affiliation{London Center for Nanotechnology and Department of Physics and
Astronomy, University College London, Gower Street, London WC1E 6BT, UK}
\author{B. Keimer}
\affiliation{Max-Planck-Institut f\"{u}r Festk\"{o}rperforschung, Heisenbergstrasse 1, D-70569 Stuttgart, Germany}

\begin{abstract}
 
The anisotropic complex dielectric response was  accurately extracted from spectroscopic ellipsometry measurements at phonon frequencies for the three principal crystallographic directions of an orthorhombic  ($Pbnm$) YTiO$_3$ single crystal. We identify all twenty five infrared-active phonon modes allowed by symmetry, 7B$_{1u}$, 9B$_{2u}$, and 9B$_{3u}$, polarized along the $c$-, $b$-, and $a$-axis, respectively.   From a classical dispersion analysis of the complex dielectric functions $\tilde\epsilon(\omega)$ and their inverses -1/$\tilde\epsilon(\omega)$ we define the resonant frequencies, widths, and oscillator strengths of the transverse (TO) and longitudinal (LO) phonon modes. We calculate eigenfrequencies and  eigenvectors of  B$_{1u}$, B$_{2u}$, and B$_{3u}$ normal modes and suggest assignments of the TO phonon modes observed in our ellipsometry spectra by comparing their frequencies and oscillator strengths with  those resulting from the present lattice-dynamics study.
Based on these   assignments, we estimate  dynamical effective charges of the atoms in the YTiO$_3$ lattice. We  find that, in general, the dynamical effective charges in YTiO$_3$ lattice are  typical for a family of
perovskite oxides. By contrast to a ferroelectric BaTiO$_3$, the dynamical effective charge of oxygen related to a displacement along the $c$-axis does not show the anomalously large value. At the same time, the dynamical effective
charges of Y and $ab$-plane oxygen exhibit anisotropy, indicating strong hybridization along the $a$-axis.

\end{abstract}


\date{\today}

\maketitle

\section*{I. INTRODUCTION}
The richly diverse behavior of rare-earth titanates $R$TiO$_3$ with a single $3d$ electron in the nearly degenerate Ti $t_{2g}$ orbitals is a consequence of the interplay between electron correlations, spin-orbit coupling, and
electron-lattice coupling including the Jahn Teller (JT) effect 
\cite{Imada,Mizokawa,Sawada,Keimer,Khaliullin1,Khaliullin2,Ulrich1,Akimitsu,Mochizuki,Zhou,Cwik,Solovjev,Mozhegorov,Pavarini}. Fine adjustment of the perovskite lattice in the titanates, in accordance with the principle of close packing, occurs  mainly due to the cooperative rotations of the TiO$_6$ octahedra, which are strongly dependent on the ionic radius of the rare-earth atom. The orbital structure that arises due to these distortions is determined by vibronic interactions between the lattice and the orbital subsystem. These octahedral distortions are also coupled to the magnetic ground state in $R$TiO$_3$: upon decreasing the ionic radius of $R$ from La to Y, the transition temperature of the $G$-type antiferromagnetic state observed in LaTiO$_3$ ($T_N = 150$ K) shifts to progressively lower temperatures, and the ground state ultimately becomes ferromagnetic with $T_C$ = 30 K in YTiO$_3$ \cite{Cwik,Schmitz,Greedan,Komarek}. 
Point-charge \cite{Imada,Mochizuki,Schmitz} and density functional \cite{Solovjev,Pavarini} calculations reproduce many aspects of the physical properties of the titanates, including the magnetic ground states of the end-point compounds LaTiO$_3$ and YTiO$_3$, based on the experimentally determined lattice structure. These calculations imply that the $t_{2g}$ levels  are split by $\sim$ 100-200 meV, so that the Ti orbitals are almost frozen in the experimentally relevant temperature range. However, some aspects of the magnetic \cite{Keimer,Ulrich1} and orbital \cite{Ulrich2} excitation spectra are difficult to reproduce in the framework of these rigid-orbital models. This has stimulated theories based on substantial fluctuations in the orbital sector \cite{Khaliullin1,Khaliullin2}. While such fluctuations are not apparent in the specific heat \cite{Fritsch}, recent thermal conductivity measurements of $R$TiO$_3$ have been interpreted as evidence of orbital fluctuations \cite{Goodenough}. Other recent studies of YTiO$_3$ single crystals have revealed anomalous temperature dependencies of the optical spectral weight \cite{Kovaleva1} and thermal expansion \cite{Knafo}
with an onset around 100 K, well above $T_C$, which may be manifestations of a temperature-dependent rearrangement of the $t_{2g}$ orbitals.

Because of the intimate relationship between lattice structure, orbital occupation, and magnetism, the lattice dynamics plays an important role in the unusual properties of the rare-earth titanates. In particular, one would expect anharmonic behavior and damping of the most strongly coupled modes in the spin-lattice-orbital sector due to orbital fluctuations, as well as anomalous energy shifts of phonons around the magnetic transition temperature \cite{Zhang,Saitoh}. Therefore experimental and theoretical studies of the lattice dynamics have the potential to become quantitative diagnostics of the mechanisms underlying the unusual phenomena in the rare-earth titanates. However, as little information about the phonon spectra and the assignment of the phonon modes of $R$TiO$_3$ is thus far available, this potential has not yet been exploited. 

The approach described here combines theoretical lattice-dynamics calculations with accurate spectroscopic ellipsometry measurements on well-characterized, nearly stoichiometric YTiO$_3$ single crystals of high quality. From the ellipsometry experiments we obtain the frequencies and oscillator strengths of all infrared (IR)-active TO phonon modes allowed by the $Pbnm$ symmetry. The lattice-dynamics calculations have been  performed using the GULP code \cite{Gale}. We calculate the eigenfrequencies of the Brillouin-zone center normal modes of B$_{1u}$, B$_{2u}$, and B$_{3u}$ symmetry and the corresponding eigenvector components. Based on these calculations, we suggest assignments
of the TO phonon modes observed in our ellipsometry spectra by comparing their frequencies and oscillator strengths with the calculated values. In
addition, we estimate the dynamical effective charges of the atoms in the YTiO$_3$ lattice using the calculated eigenvector components and the experimental phonon parameters of the TO frequencies and oscillator strengths. The assignment is further verified by {\it ab initio} lattice-dynamics calculations \cite{Capogna} within the density functional approach using the VASP software package \cite{VASP,PHONON}.
The present study provides a coherent picture of the lattice vibrations in YTiO$_3$, consistent with both experimental data and with full electronic structure calculations.

\section*{II. CRYSTAL STRUCTURE AND $\Gamma $-POINT PHONONS IN
$\mathrm{YTiO_3}$}

YTiO$_3$ exhibits an orthorhombic structure of a GdFeO$_3$ type described
by the $Pbnm$ space group ($Pnma$ $D^{16}_{2h}$, No. 62 in the International Tables for X-ray Crystallography \cite{Tables}, with nonconventional coordinate axes orientation, $Pnma (a,b,c)$ $\Longleftrightarrow$ $Pbnm (b,c,a)$), with 4 f.u./unit cell (see Fig. 1, on the right side). This orthorhombic structure can be regarded   as a distorted modification of the  simple cubic ABO$_3$ perovskite structure with space group $Pm\bar3m$ (see Fig. 1, on the left side). The orthorhombic distortions can be attributed to the common atomic size mismatch: the sum of the ionic radii in the Ti-O$_2$ layer, $r_{Ti} + r_O$, does not match that of the Y-O layer, $(r_Y+r_O)/\sqrt2$, in the right way for a stable cubic structure. These orthorhombic distortions lead first to  the oxygen ion displacements, resulting from the two consequent and coordinated rotations of the TiO$_6$ octahedra, described by $a^-a^-b^+$ in the Glazer's notations \cite{Glazer}, and second to a modification of the Y-O and Y-Ti coordinations. In addition, there are JT-type distortions, resulting in a slight extension or compression of the Ti-O bond pairs, and the ``scissors"-type distortions of the TiO$_6$ octahedron \cite{Cwik,Schmitz,Mozhegorov}.
The Wyckoff positions of the atoms and their site symmetries in the orthorhombic unit cell are listed in Table 1. The irreducible representations (in the $Pbnm$ notation) corresponding to various atomic sites in the orthorhombic structure that follow from the character tables of the point groups are presented in the right column of Table 1. The total numbers of the modes are grouped according to their optical activity. Among the total number of 60 $\Gamma$-point phonons, 24 (7A$_g$+7B$_{1g}$+5B$_{2g}$+5B$_{3g}$) are Raman-active modes, 25 (7B$_{1u}$+9B$_{2u}$+9B$_{3u}$) are IR-active modes, 8 (A$_u$) are silent modes, and 3 (B$_{1u}$+B$_{2u}$+B$_{3u}$) are acoustic modes.    

\section*{III. SAMPLES AND EXPERIMENTAL APPROACH}

\subsection{Crystal growth and characterization}

It has to be emphasized that YTiO$_{3+\delta}$ single crystals are always slightly off-stoichiometric, having an excess of oxygen above the exact stoichiometry formula. As to our knowledge, a minimal level record of $\delta$ = 0.009
belongs to single crystals of YTiO$_{3+\delta}$ grown by Czochralski method
\cite{MacLean}. In this study, single crystals of YTiO$_3$ were grown by the floating zone method in a reducing atmosphere (Ar/H$_2$ = 50/50). More details are given in Ref. \cite{Kovaleva1}. The oxygen excess in our samples
was measured by the Differential Thermal Analysis and Thermal Gravimetry
Analysis and estimated at a level less than $\delta$ = 0.013. 
The same level of the oxygen off-stoichiometry $\delta$ = 0.01 was determined in the single crystals of YTiO$_{3+\delta}$, grown by the floating zone method in a fairly strong reducing condition (Ar/H$_2$ = 70/30) by Okimoto {\it et al.} \cite{Okimoto}. The optical conductivity spectra measured on our crystals and those grown by  Okimoto {\it et al.} are very well agree with
each other in a wide spectral range, from the phonon frequencies up to
the deep UV frequencies (see Fig. 7 in Ref. \cite{Kovaleva1} and Fig. 2 in
Ref. \cite{Okimoto}). Therefore, we conclude that our single crystals of YTiO$_3$ grown by the floating zone method are of the same quality, verified by the optical probe. Moreover, the Ti$^{3+}$ ions in the YTiO$_3$ lattice are unstable in oxygen atmosphere, and fully oxidize to Ti$^{4+}$ above 300 $^\circ$C in flowing oxygen (see Fig. 2 in Ref. \cite{Kovaleva1}). This may lead to the aging effects in air, associated with the formation of the titanium oxides with the valence of Ti increased above 3+ and   simultaneous incorporation of extra oxygen (interstitials). The aging effects and the oxygen off-stoichiometry
in YTiO$_{3+\delta}$ crystals have a deep impact on the optical conductivity spectra and their temperature dependencies \cite{Kovaleva1,Goessling}. If the samples are kept in air for extending periods of time, the weak satellites
can appear in the phonon modes, as arising from the coupling with the local oxygen modes associated with oxygen interstitials, as in the case of the
$ab$ surface, reported in Ref. \cite{Kovaleva1}. Therefore, an accurate study of the optical properties in YTiO$_{3+\delta}$ crystals should be always referred to the oxygen off-stoichiometry $\delta$.   

The samples were further characterized by magnetometry using a superconducting
quantum interference device. For the almost stoichiometric YTiO$_{3+0.013}$ sample studied here, we estimate a Curie temperature of  T$_C$ = 30 K, an
inflection point in the temperature-dependent magnetization is observed at
27 K. The determined saturated magnetic moment in the easy direction at 7 T is  $\mu^c$ = 0.84 $\mu_B$, whereas $\mu^b$ = 0.82 $\mu_B$ in the hard
direction \cite{Kovaleva1}. These our data are in good agreement with the
magnetization data on the nearly oxygen stoichiometric YTiO$_3$ single crystals,
which referred to T$_C$ of 29 K \cite{Garret} and 30 K \cite{Tsubota,Tsuji}.

At the same time x-ray diffraction measurements show that the YTiO$_3$ single crystals used for our experiments are untwinned, and posses an excellent
crystal quality with mosaicity less than 0.03$^\circ$. At room temperature, the lattice parameters are $a$ = 5.331(3), $b$ = 5.672(4), and $c$ = 7.602(6) $\AA$, slightly different from those of Czochralski-grown single crystals reported by Maclean {\it et al.} \cite{MacLean}. The samples were aligned along the principal axes and cut in the form of a parallelepiped with dimensions $\sim 3 \times 3 \times 3$ mm$^3$.   

\subsection{Ellipsometry approach}

For optical measurements the sample surfaces were polished to optical grade. The technique of ellipsometry provides significant advantages over conventional reflection methods in that (i) it is self-normalizing and does not require reference measurements and (ii) $\epsilon _{1}(\omega )$ and $\epsilon _{2}(\omega )$ are obtained directly without a Kramers-Kronig transformation
\cite{Henn,Bernhard,Handbook}. For ellipsometry measurements at phonon frequencies 50-700 cm$^{-1}$ (0.006-0.087 eV) we used the home-built ellipsometer in combination  with a fast-Fourier transform interferometer (Bruker IFS 66v/S) at the infrared beam line of the ANKA synchrotron at the Forschungszentrum Karlsruhe, Germany. The high brilliance of the synchrotron light source due to the small beam divergence enabled us to perform very accurate ellipsometric measurements in the far-IR range with incidence angles ranging from 65$^\circ$ to 85$^\circ$. For low-temperature measurements the sample was mounted on the cold finger of a helium flow cryostat. With only a single angle of incidence, the raw experimental data  are represented by real values of the ellipsometric angles, $\Psi$ and $\Delta$, for any wave number. These values are defined through the complex Fresnel reflection coefficients for light polarized parallel ($r_p$) and perpendicular ($r_s$) to the plane of incidence
\begin{eqnarray}
{\rm tan} \ \Psi e^{i\Delta}=\frac{r_p}{r_s}.
\end{eqnarray}
To determine the complex dielectric response $\tilde\epsilon_a(\omega)$, $\tilde\epsilon_b(\omega)$, and $\tilde\epsilon_c(\omega)$ of the YTiO$_3$ crystal, we measured four high-symmetry
orientations of the $ac$ and $bc$ surfaces, with $a$ or $b$ axes aligned either parallel or perpendicular to the plane of incidence of the light, respectively. Note here that  we have used an overdetermined data set to check the consistency of our results. A nonlinear fitting procedure
has been applied to extract point-by-point the complex dielectric response
throughout the covered spectral range \cite{Humlicek}. In the following,
we present the complex dielectric response $\tilde\epsilon(\omega)$ extracted from the raw ellipsometry spectra, $\Psi(\omega)$ and $\Delta(\omega)$, according to this approach.             
             
\section*{IV. RESULTS AND DISCUSSION}

\subsection*{A. Infrared spectra of YTiO$_3$}

Figures 2-4 show the dielectric functions $\tilde\epsilon(\omega)$ and their inverses $-1/\tilde\epsilon(\omega)$ for the three principal crystallographic axes of the YTiO$_3$ crystal, measured with an angle of incidence of 75$^\circ$ at T = 25 K. As discussed in Sec. II, a factor-group analysis for the orthorhombic crystal structure of YTiO$_3$ (space group $Pbnm$, $D^{16}_{2h}$) with 4 f.u./unit cell yields a total number of 25 (7B$_{1u}$+9B$_{2u}$+9B$_{3u}$) IR-active modes. Our polarized ellipsometry  measurements allowed us to observe all symmetry-allowed  IR-active phonon modes. One can easily identify six phonon modes in the polarization along the $c$ axis (vibrations of B$_{1u}$ symmetry), eight phonon modes in  the $b$ axis (B$_{2u}$), and
eight phonon modes in the $a$ axis (B$_{3u}$). In addition, we were able to identify very weak features in our raw ellipsometry spectra, $\Psi(\omega)$ and $\Delta(\omega)$, on the high-frequency side in each polarization (not shown), which we associate with B$_{1u}$(7), $B_{2u}$(9), and B$_{3u}$(9) symmetry modes. Using the classical dispersion analysis we fit the complex dielectric functions $\tilde\epsilon_a(\omega)$, $\tilde\epsilon_b(\omega)$, and $\tilde\epsilon_c(\omega)$ (and their inverses independently) with a set of Lorenzian oscillators
\begin{eqnarray}
\tilde\epsilon(\omega) = \epsilon_{\infty}
+ \sum_j \frac{S_j \omega^2_j}{\omega_j^2-\omega^2-i\omega\gamma_j},
\end{eqnarray}
where $\omega_j$, $\gamma_j$, and $S_j$ are the resonant frequency, width, and dimensionless oscillator strength of the $j$th oscillator, and $\epsilon_{\infty}$ is the core contribution from the dielectric function \cite{note}. The resulting parameters determine $\omega_{{\rm TO}j}$, $\gamma_{{\rm TO}j}$, and $S_j$ of the transverse TO phonon modes (and $\omega_{{\rm LO}j}$ and $\gamma_{{\rm LO}j}$ of the longitudinal LO phonon modes). We determine  the anisotropic static dielectric constants $\epsilon_0$ from our fit at 50 cm$^{-1}$ (these values are given in Table IV), and estimate the anisotropic high-frequency dielectric constants defined by the relation
$\epsilon_0 = \epsilon^*_{\infty}{\prod}_m \frac{\omega^2_{LO,m}}{\omega^2_{TO,m}}$. In Table II we list all Lorentzian parameters, together with $\epsilon_{\infty}$ and $\epsilon^*_{\infty}$, which show slightly different values, according
to the accuracy of the present study.      
 
\subsection*{B. Theoretical calculations}

\subsubsection{\it {Shell model approximation and potential parameters}}

To determine the eigenfrequencies and eigenvectors  of the
normal modes  and to make an assignment  in the experimental phonon spectra,
we have performed lattice-dynamics calculations for the orthorhombic YTiO$_3$ structure in the $Pbnm$ (D$^{16}_{2h}$) space group in the framework of a shell model. The shell model approach has been successfully applied earlier to study various physical properties and lattice dynamics of transition metal oxides, including high-$T_c$ superconductors \cite{Prade,Islam,Kovaleva2} and LaMnO$_3$ and YMnO$_3$ \cite {Iliev,Smirnova,Kovaleva3}. Our calculations were performed in the GULP code \cite{Gale}, using the lattice constants and fractional coordinates of the atoms  reported for the Czochralski-grown single crystals by Maclean {\it et al.} \cite{MacLean}.
In the shell model approximation, the lattice is considered as an assembly of polarizable ions, represented by massive point cores and massless shells coupled by isotropic harmonic forces defined by the spring constants. The interaction includes contributions from Coulomb and short-range interactions.
The short-range potentials used for the shell-shell interactions are of the Buckingham form
\begin{eqnarray}
V_{ij} = A_{ij} exp {(-r/\rho_{ij})} - C_{ij}/r^6.
\end{eqnarray}
The  O-O (shell-shell) interactions are taken as those of typical oxides \cite{Catlow,Popov}. We adopt the formal ionic charges here.
The forces acting on the atoms in the lattice were minimized by varying the Buckingham potential parameters, spring constants, as well as the core and shell charges. The crystalline structure was allowed to relax to equilibrium conditions under the symmetry ({\it Pbnm}) restrictions, the cores and shells were allowed to move separately. The resulting Buckingham potential parameters, shell charges (Y)  and force constants (k) are presented in Table III. The criterion for a successful fit was good agreement with the structural parameters \cite{MacLean}, provided that approximately zero forces are applied for all atoms, and with the experimental anisotropic static $\epsilon_0$ and high-frequency $\epsilon_{\infty}$ dielectric constants, determined in the present spectroscopic ellipsometry study (as Table IV illustrates).   

\subsubsection{Calculated $\Gamma$-point IR-active modes in YTiO$_3$}

A factor group analysis of the zone-center phonon modes in the orthorhombic  YTiO$_3$ ($Pbnm$) structure yields 25 IR-active phonon modes, which have their progenitors in the Brillouin-zone of the cubic perovskite ABO$_3$ ($Pm\bar3m$) structure \cite{Couzi}. Among them, there are nine phonon modes, 3B$_{1u}$ + 3B$_{2u}$ + 3B$_{3u}$, emanating from the three IR-active F$_{1u}$ cubic triplets. These  are commonly referred to the $external$ mode (in which the BO$_6$ octahedron vibrates against the A atoms), the $bending$ mode (B-O-B bond-angle modulation), and the $stretching$ mode (B-O bond-length modulation). In addition, two IR-active modes of B$_{2u}$ and B$_{3u}$ symmetry and one silent mode of A$_u$ symmetry stem from the cubic F$_{2u}$ silent triplet. The rest orthorhombic IR modes arise from $X$, $M$, and $R$ points of the cubic Brillouin-zone, and become active due to Brillouin-zone folding in the orthorhombic phase \cite{Couzi}. However, in this study we  are not aimed at a detailed description of the orthorhombic phonon eigenvectors in terms of the corresponding cubic eigenvectors. Here we present results of our lattice-dynamics shell-model (SM) calculations and make an assignment of the IR-active phonon modes $\omega_{{\rm TO}j}$ observed in the polarized ellipsometry spectra $\tilde\epsilon(\omega)$
in the YTiO$_3$ single crystal (see Figs. 2-4 and Table II).

We tentatively divide the observed phonon modes into three groups centered at about the characteristic frequencies of $external$ (170 cm$^{-1}$), $bending$ (340 cm$^{-1}$), and $stretching$ modes (540 cm$^{-1}$) of the rare-earth perovskite oxides. Correspondingly, the eigenvector patterns for all normal modes of each symmetry B$_{1u}$, B$_{2u}$, and B$_{3u}$, as derived from the SM lattice-dynamics calculations, are presented in Figs. 5-7. Here only major components of the eigenvectors (the whole list is given in Table V) along the principal crystallographic directions are displayed for 20 atoms in the unit cell, of different symmetries and in all inequivalent positions. At each pattern the calculated eigenfrequencies are indicated. As one can notice from Table VI, the calculated eigenfrequencies generally well agree with the experimental TO phonon frequencies $\omega_{{\rm TO}j}$. The character of the B$_{1u}$, B$_{2u}$, and B$_{3u}$ modes is determined according to
the dominant contribution from the collective vibrations of Y, Ti, O1, or O2 atoms, polarized along the $c$-, $b$-, and $a$-axis, respectively. However,
mixing due to out-of-phase vibrations  of ions of equal symmetry in inequivalent positions often does not allow the determination of a unique character for a given set of equal symmetry modes.

At low frequencies, near the characteristic frequency of the cubic perovskite $external$ mode, one can identify three modes, B$_{1u}$(1) : 167 cm$^{-1}$, B$_{2u}$(2) : 240 cm$^{-1}$, and B$_{3u}$(2) : 209 cm$^{-1}$, which exhibit very high oscillator strengths in the experimental IR phonon spectra, in agreement with the SM calculations (see Table VI). The eigenvectors of the B$_{1u}$(1) mode show that the oscillator strength of this mode is determined by vibrations of the TiO$_6$ octahedra against the Y atoms along the $c$-axis. The eigenvectors of the B$_{2u}$(2) and B$_{3u}$(2) modes show that these normal modes involve dominant contribution from the collective Y-Ti vibrations, where Y and Ti atoms move in the opposite directions, polarized in accordance with the mode symmetries along the $b$- and $a$-axis, respectively. The eigenvectors of the B$_{2u}$(1) : 139 cm$^{-1}$ and B$_{3u}$(1) : 154 cm$^{-1}$  modes indicate that these modes are primarily due to out-of-phase vibrations of different Y atoms and out-of-phase vibrations of different Ti atoms, polarized
respectively along the $a$- and $b$-axis, not coincident with the mode symmetries.
In agreement with the SM calculations, the observed phonon modes B$_{2u}$(1) and B$_{3u}$(1) exhibit very small oscillator strengths. Another weak mode is the $c$-axis B$_{1u}$(2) : 204 cm$^{-1}$, which is mainly due to $b$-axis in-phase vibrations of  Ti atoms in one plane and out-of-phase vibrations in different planes.

At medium phonon frequencies near the cubic perovskite $bending$ mode, the strong $c$-axis polarized modes B$_{1u}$(3) : 324 cm$^{-1}$ and B$_{1u}$(5) : 387 cm$^{-1}$ are mainly due to Ti-O2-Ti $bending$ vibrations (the B$_{1u}$(3) mode has small admixture of O1 character due to the movement of O1 atoms in-phase with Ti atoms). The $b$-axis polarized modes B$_{2u}$(3) : 273 cm$^{-1}$ and B$_{2u}$(7) : 461 cm$^{-1}$ involve collective vibrations of O2 and Y atoms moving in the opposite directions. In turn, the  strong $a$-axis polarized modes B$_{3u}$(3) : 316 cm$^{-1}$ and B$_{3u}$(6) : 426 cm$^{-1}$ also involve collective vibrations of O2 atoms (the B$_{3u}$(3) mode has in addition small contribution due to in-phase vibrations of O1 atoms and out-of-phase vibrations of Y atoms). The character of the mode B$_{2u}$(4) : 308 cm$^{-1}$ is defined by the dominant contribution from O1 vibrations, moving out-of-phase with O2 atoms. The modes B$_{2u}$(5) : 364 cm$^{-1}$ and B$_{3u}$(5) : 344 cm$^{-1}$ can be assigned as Ti-O1-Ti $bending$ modes. Other modes in this range B$_{2u}$(6) : 380 cm$^{-1}$ and B$_{3u}$(7) : 516 cm$^{-1}$ are also determined by O1 collective vibrations, polarized according to the mode symmetries.  We note
here that the two $b$-axis modes B$_{2u}$(5) : 364 cm$^{-1}$ and B$_{2u}$(6) : 380 cm$^{-1}$ having very close frequencies could be reversed in the assignment, as it is explained below. The $c$-axis B$_{1u}$(4) : 349 cm$^{-1}$ has
very small oscillator strength as it can be assigned to the $a$-axis out-of-phase
vibrations of Ti atoms.

In the third group, the strong modes B$_{1u}$(6) : 545 cm$^{-1}$,
B$_{2u}$(8) : 528 cm$^{-1}$, and B$_{3u}$(8) : 554 cm$^{-1}$ are observed in the experimental IR phonon spectra. Therefore, these IR-active phonon modes of orthorhombic YTiO$_3$ should be most closely related to the cubic perovskite $stretching$ modes. Indeed, the eigenvector patterns
show that the B$_{1u}$(6) mode is due to the $c$-axis  Ti-O1 vibrations resulting
in the Ti-O1 bond-length modulation, and the oscillator strengths of the B$_{2u}$(8) and B$_{3u}$(8) normal modes are determined by the $b$- and $a$-axis Ti-O2 collective out-of-phase vibrations. In addition, we were able to identify very weak features in our ellipsometry spectra on the high-frequency side in each polarization, which we  associate with the B$_{1u}$(7) : 595 $cm^{-1}$, $B_{2u}$(9) : 577 cm$^{-1}$, and B$_{3u}$(9) : 576 cm$^{-1}$ symmetry modes. We note that for the B$_{1u}$(6) and B$_{1u}$(7) modes there is a good agreement between the frequencies and oscillator strengths, predicted from the calculations and those observed (see Table VI). The B$_{1u}$(7) normal mode is weak, because it is mostly due to the different symmetric $ab$-polarized
out-of-phase oscillations of Ti and O2 atoms. However, our SM calculations predict that the strong in-plane modes of $B_{2u}$ and $B_{3u}$ symmetry appear at the high-frequency boundary of the phonon spectra. In this case, in our assignment we relate the modes of similar oscillator strengths. In agreement with our SM calculations the B$_{2u}$(9) and B$_{3u}$(9) normal modes are largely due to the symmetric $a$- and $b$-polarized out-of-phase vibrations of Ti and O1 atoms, respectively. Regarding the symmetry of these modes, their oscillator strengths is expected to be small, in consistency with our experimental observations.

\subsubsection{Comparison with first-principles results}

In order to independently support the assignment of the IR-active phonon modes in YTiO$_3$, we have calculated the $\Gamma$-point normal modes  by means of the {\it ab initio} approach within the density functional theory using the VASP software package \cite{VASP,PHONON}. A detailed description of these calculations will be given elsewhere \cite{Capogna}. Here we only mention the eigenfrequencies of the normal modes resulting from the {\it ab initio} calculations:  $a$-axis 9 B$_{3u}$ normal modes
are calculated at the frequencies 137, 175, 298, 312, 330, 379, 391, 473, and 519 cm$^{-1}$;  $b$-axis 9 B$_{2u}$ normal modes  at 125, 211, 250, 291, 337, 359, 440, 485, and 500 cm$^{-1}$; and  $c$-axis 7 B$_{1u}$ normal modes
at  144, 182, 289, 336, 352, 463, and 517 cm$^{-1}$. In general, the frequencies
and eigenvectors of the normal modes predicted by the {\it ab initio} lattice-dynamics calculations are in a reasonable agreement with the results of the shell model lattice-dynamics calculations reported in detail here, especially for the low-frequency and high-frequency modes, near the characteristic frequencies of the cubic perovskite $external$ and $stretching$ modes, respectively. The main characters of the modes near the characteristic cubic perovskite $bending$ mode (where the phonon frequencies are densely spaced  and therefore the results are very sensitive to details) are also in satisfactory agreement with the shell model lattice-dynamics calculations. For example, the assignments  of the $b$-axis modes B$_{2u}$(3) : 273 cm$^{-1}$ and B$_{2u}$(4) : 308 cm$^{-1}$, the $a$-axis mode B$_{3u}$(6) : 426 cm$^{-1}$ and the $c$-axis mode B$_{1u}$(3) : 324 cm$^{-1}$ are well reproduced.
    
\subsubsection{Effective charges in YTi$O_3$}

The {\it static} effective charges $Z_i e$ related to the bonding ionicity can be defined for a ternary compound as follows \cite{Scott,Tajima}
\begin{eqnarray}
\sum_j[\omega^2_{{\rm LO}j}-\omega^2_{{\rm TO}j}] = \frac{4\pi}{V_c}\sum_i (Z_i e)^2/M_i,
\end{eqnarray}
where $j$ denotes the phonon mode with TO frequency $\omega_{{\rm TO}j}$ and LO frequency $\omega_{{\rm LO}j}$, $V_c$ is   the unit cell volume, $M_i$ corresponds to the mass of the $i$th atom. The static effective charges of oxygen $Z_O e$ can be estimated rather accurately from this relation, as the contribution of the terms related to oxygen is dominant on the right hand side of Eq. (4) due to the small mass of oxygen as compared with the masses of other constituents. Using the frequencies   $\omega_{{\rm TO}j}$ and $\omega_{{\rm LO}j}$ determined from the ellipsometry
measurements listed in Table II,  we have found that the static effective charges of oxygen in orthorhombic YTiO$_3$ are nearly isotropic: $Z_{O_{\parallel a}}$ = -1.14 $\mid e \mid$, $Z_{O_{\parallel b}}$ = -1.17
$\mid e \mid$, and $Z_{O_{\parallel c}}$ = -1.20 $\mid e \mid$. These values
are typical for the entire family of perovskite oxides.

The Born effective charge of an ion  represents the dynamical contribution to the effective charge. It is controlled by long-range Coulomb interactions. The {\it dynamical} effective charge is represented by a tensor, since it is defined by the change of macroscopic polarization created in direction $\beta$, induced by the periodic displacement $\tau_{\kappa,\alpha}$
of the sublattice of atoms $\kappa$ in direction $\alpha$ at the linear order,
times the unit cell volume
\begin{eqnarray}
Q^*_{\kappa,\alpha\beta}=\ V_c\frac{\partial \mathcal{P}_{\beta}}{\partial
\tau_{\kappa,\alpha}}.   
\end{eqnarray}
For the case of ABO$_3$ compounds, the values of the Born effective charges
deviate substantially from the static effective charges. The {\it ab initio} calculations report anomalously large values of dynamical effective charges
for Ti and O ions in ferroelectric perovskites KNbO$_3$ and BaTiO$_3$ \cite{Resta,Ghosez}. The charges of Ba and Ti atoms in the orthorhombic structure Q$^*_{Ba}$ and Q$^*_{Ti}$ of BaTiO$_3$ are nearly isotropic and vary, respectively, in the range from +2.72 to +2.77 and from +5.59 to +6.8. In turn, for oxygen, the effective charges exhibit predominantly uniaxially anisotropic character, where the values of Q$^*_{O_{\parallel}}$ vary from -1.91 to -2.04 and Q$^*_{O_{\perp}}$ vary from -4.89 to -5.45, and refer, respectively, to a displacement of oxygen
in the $ab$-plane and along the $c$-axis \cite{Ghosez}.

The oscillator strengths $S_j$ of IR-active phonon modes are related to the dipole moment arising from the displacements of ions involved in a vibration
\begin{eqnarray}
S_j = \frac{4\pi}{V_c}\frac{[\sum_iQ^*_{t_i}u_{ij}]^2}{\omega^2_{{\rm TO}j}\sum_i
M_iu^2_{ij}}.
\end{eqnarray}
Here $Q^*_{t_i}$ corresponds to  the dynamical effective charge
of the $i$th atom vibrating in $j$ mode with the TO frequency $\omega_{{\rm TO}j}$, and $u_{ij}$ represents the displacements related to the eigenvector $\chi_j$ by $u_{ij} =\chi_{ij}/\sqrt{M_i}$.  In Table VI we present calculated
squares of net dipole moments $[\sum_iQ_iu_{ij}]^2$ arising from the displacements of all ions involved in the different symmetry IR-active vibrations, using the formal ionic charges that were adopted in our shell model. These values are juxtaposed with the corresponding values estimated from the experimental phonon parameters of $\omega_{\rm{TO_j}}$ and oscillator strengths $S_j$ given in Table II. As one can see, the qualitative agreement between the model and the experiment is quite good in the sense that the model correctly predicts the strongest modes in the phonon spectra and the weak modes are also predicted to be weak. However, some of the  modes, for example, two strong $b$-axis modes B$_{2u}$(3) : 273 cm$^{-1}$ and B$_{2u}$(6) : 380 cm$^{-1}$, two strong $a$-axis modes B$_{3u}$(2) : 209 cm$^{-1}$, B$_{3u}$(6) : 426 cm$^{-1}$, and one  weak mode B$_{3u}$(7) : 516 cm$^{-1}$, and the strong $c$-axis mode B$_{1u}$(3) : 324 cm$^{-1}$ show largest discrepancy  between the experimental values and calculated in the framework of the adopted shell model. 

An improved description of the experimental oscillator strengths of all phonon modes compared to those with formal ionic charges can be achieved by introducing
Born effective charges. These effective charges can be extracted by inversion the linear equations Eq. (6) by making use of the eigenvectors, providing the  experimental phonon parameters of $\omega_{\rm{TO_j}}$ and $S_j$ are available. We note here that for each polarization together with the electric charge neutrality condition,  we have  an over determined data set for the four unknown effective charges. However, the inversion of Eq. (6) results in large errors if it is applied for the  modes  which have eigenvector components  close to zero, in particular, for the weak modes (see Tables V and VI), as the determinant of the matrix is then close to zero. We apply Eq. (6) to the three strong modes in each polarization  and determine the corresponding dynamical effective charges, which are listed in Table VII. In reasonable agreement with the effective charges of Ti atoms in the orthorhombic structure of BaTiO$_3$, with the nominal valence of Ti atoms +4 \cite{Ghosez}, we have
found that in YTiO$_3$ the  dynamical effective charges of Q$^*_{Ti}$ are nearly isotropic, varying in the range from +3.9 to +5.00. At the same time, the effective charges of Y and O2 atoms exhibit $a$-axis anisotropy: Q$^*_{Y{\parallel}a}$=+1.5 and Q$^*_{O_2{\parallel}a}$=-1.9, compared to the corresponding effective charges Q$^*_{Y{\parallel}b,c}$ of +3.4 \(\div\) +3.9 and Q$^*_{O_{1,2}{\parallel}b}$ of -2.5 \(\div\) -2.7. The dynamical effective charges of oxygen Q$^*_{O_{1,2}{\perp}}$,
related to a displacement along the c-axis, vary in the range from -2.3 to
-2.8 in YTiO$_3$ and do not show anomalously large  values like in ferroelectric BaTiO$_3$ \cite{Ghosez}. In Table VI the calculated values of the $[\sum_i Q^*_iu_{ij}]^2$ are juxtaposed with the corresponding experimental values and those obtained from the calculations with formal ionic charges. One can notice that, in general, the agreement with the experimental values is improved (except for some of  the weak modes,  which are less accurately predicted). The description with the effective charges give also an indication that  two $b$-axis modes B$_{2u}$(5) : 364 cm$^{-1}$ and B$_{2u}$(6) : 380 cm$^{-1}$ having very close frequencies should be reversed in the assignment. The agreement for the $a$-axis modes B$_{3u}$(6) : 426 cm$^{-1}$ and B$_{3u}$(7) : 516 cm$^{-1}$ is not substantially improved by introducing Born effective charges. A possible reason could be a strong dependence of the mode B$_{3u}$(7) on strain in the real crystal structure. As a result, most of its oscillator strength might be transferred to the strongest $a$-axis mode  B$_{3u}$(6). We note that the $b$-axis mode B$_{2u}$(3) : 273 cm$^{-1}$ is less successfully predicted, however, the character of this mode is well reproduced by
the present $ab \ initio$ calculations \cite{Capogna}.
        
\section*{V. CONCLUSIONS}

In summary, we present optical phonon spectra of the orthorhombic YTiO$_3$ single crystal as obtained by spectroscopic ellipsometry and suggest the assignments of all symmetry-allowed zone-center IR-active phonon modes by comparing them with the results of the shell model calculations. We note that the shell model elaborated in this work may possess certain advantages, as the parameters of the model were fitted not only to the orthorhombic crystal structure, but also to the experimental anisotropic static $\epsilon_0$
and high-frequency $\epsilon_\infty$ dielectric constants, determined in
the present ellipsometry study. Consequently, the calculated dipole moments (or oscillator strengths) associated with displacements of the atoms involved in the IR-active  normal modes in YTiO$_3$ are in good agreement with the experimental data. The normal modes predicted in the framework of the present shell model are in good agreement with the {\it ab initio} lattice-dynamics calculations \cite{Capogna}. Thus, our  study gives us a coherent picture of the lattice vibrations in YTiO$_3$, consistent with both experimental data and with full electronic structure calculations. We have not considered the influence of magnetic interactions on the lattice dynamics, but as the ferromagnetic coupling in YTiO$_3$ is weak it should not affect our results substantially.

In order to describe the lattice dynamics of complex oxides like the rare-earth  titanates, one  can easily imagine a need to go beyond the shell model combined with interatomic potentials. However, the shell model itself is a general model within the dipole and harmonic approximations \cite{Cochran}, and there have been wide-ranging and successful calculations for simple binary oxides \cite{Stoneham1}, including predictions of surface structure and vibrations, off-center ions, charge transfer transitions, and defect equilibria for non-stoichiometric oxides. What was unexpected is the extent to which the oxygen-oxygen potentials seem to be transferrable to the more complex oxides. One might have expected the Jahn-Teller effect or charge transfers and lack of ionicity to lead to radical changes. This seems not to be the case: we find the O-O potentials based on MgO and similar binary oxides essentially the same for LaMnO$_3$, for Bi-based oxide superconductors, and for YTiO$_3$. Some of the reasons for this are discussed in relation to oxide superconductors in Ref. \cite{Stoneham2}.
As a result, the shell model combined with interatomic potentials continues
to offer a powerful route to understanding experimental data for complex oxides.

The comprehensive study described here provides a solid foundation for further quantitative exploration of the interplay between spin, orbital and lattice degrees of freedom in the titanates. We found that in YTiO$_3$ the effective charges of Ti are nearly isotropic and comparable to the effective charges of Ti in other perovskite compounds. At the same time the effective charges of Y and O2 atoms exhibit $a$-axis anisotropy: Q$^*_{Y{\parallel}a}$=+1.5 and Q$^*_{O_2{\parallel}a}$=-1.9, compared to the corresponding effective charges Q$^*_{Y{\parallel}b,c}$
(+3.4 \(\div\) +3.9) and Q$^*_{O_{1,2}{\parallel}b}$ (-2.5 \(\div\) -2.7).
As the effective charges are a sensitive tool for analyzing dynamic changes of orbital hybridization, we suggest strong hybridization effects between Y and $ab$-plane oxygen along the $a$-axis. Hybridization of Y $5d$ and O $2p$ states has indeed been predicted by density functional calculations \cite{Pavarini}. Then, according to our assignment, the modes B$_{3u}$(3) : 316 cm$^{-1}$ and B$_{3u}$(6) : 426 cm$^{-1}$, involving O2 and Y displacements along the $a$ axis, will probably exhibit anomalous temperature dependencies. Some of the Raman-active modes related to the ``scissors"-type distortions of the TiO$_6$ octahedra, which are especially important in YTiO$_3$ \cite{Mozhegorov}, might also exhibit anomalous temperature dependence in the vicinity of $T_C$. As in the real structure of YTiO$_3$ these distortions are most pronounced in the $bc$-plane, we expect that also some of the IR-active modes involving O1 oxygen displacements along the $b$ axis and O2 oxygen displacements along the $c$ axis will exhibit anomalous behavior around the ferromagnetic transition temperature. According to our assignment, the possible candidates are B$_{2u}$(4) : 308 cm$^{-1}$, B$_{2u}$(6) : 380 cm$^{-1}$, B$_{1u}$(3) : 324 cm$^{-1}$, and B$_{1u}$(5): 387 cm$^{-1}$.

\section*{VI. ACKNOWLEDGMENTS}

The authors thank J. Gale for making available General Utility Lattice
Program (GULP) used in the present calculations.
We are grateful to R. Evarestov,  E. Kotomin and O. Dolgov for fruitful discussions.
We acknowledge Y.-L. Mathis at ANKA for his support
during the ellipsometry measurements.

\newpage

\begin{table*}[tbp]
\caption{Wyckoff notations, atomic site symmetries, and irreducible representations
for the atoms in orthorhombic {\it Pbnm} (D$^{16}_{2h}$) YTiO$_3$.}
\begin{ruledtabular}
\begin{tabular}{ccllll}   
Atom & Wyckoff   & Site     & Irreducible  representation \\     
     & notation  & symmetry &             &                \\ \colrule
Y    & 4(c) & C$^{xy}_s$     & 2 A$_g$+A$_u$ + 2B$_{1g}$ + B$_{1u}$ + B$_{2g}$ + 2B$_{2u}$ + B$_{3g}$
+ 2B$_{3u}$ \\
Ti   & 4(b) & C$_i$          & 3 A$_u$  + 3B$_{1u}$
+ 3B$_{2u}$ + 3B$_{3u}$ \\
O1   & 4(c) & C$^{xy}_s$     & 2A$_g$ + A$_u$
 + 2B$_{1g}$  + B$_{1u}$ + B$_{2g}$ + 2B$_{2u}$ + B$_{3g}$
+ 2B$_{3u}$ \\
O2   & 8(d) & C$_1$ & 3A$_g$ + 3A$_u$ + 3B$_{1g}$ + 3B$_{1u}$ + 3B$_{2g}$ + 3B$_{2u}$ + 3B$_{3g}$ + 3B$_{3u}$\\     
\\
\colrule
&&& 7A$_g$ + 8A$_u$ + 7B$_{1g}$ + 8B$_{1u}$ + 5B$_{2g}$ + 10B$_{2u}$
+ 5B$_{3g}$ + 10B$_{3u}$\\
&&&$\Gamma_{IR}$ = 7B$_{1u}$ + 9B$_{2u}$ + 9B$_{3u}$ \\
&&&$\Gamma_{Raman}$ = 7A$_g$ + 7B$_{1g}$ + 5B$_{2g}$ + 5B$_{3g}$ \\
&&&$\Gamma_{silent}$ = 8A$_u$ \\
&&&$\Gamma_{acoustic}$ = B$_{1u}$ + B$_{2u}$ + B$_{3u}$
\end{tabular}
\end{ruledtabular}
\end{table*}

\begin{table*}[tbp]
\caption{Results of the dispersion analysis based on the three-parameter
Lorenz model: frequencies and damping parameters $\omega_{{\rm TO}j}$ and $\gamma_{{\rm TO}j}$ (in cm$^{-1}$), and dimensionless oscillator strengths $S_j$ of the TO phonon modes in the the polarized $\tilde\epsilon(\omega)$ spectra of the YTiO$_3$ single crystal (T = 25 K), $\epsilon_\infty$ is the core contribution; frequencies and damping parameters $\omega_{{\rm LO}j}$ and $\gamma_{{\rm LO}j}$ (in cm$^{-1}$) of the LO  phonon modes in the the -1/$\tilde\epsilon(\omega)$ spectra. Definition of $\epsilon^*_\infty$ is explained in the text.}
\begin{ruledtabular}
\begin{tabular}{lllllllllllllll}
$a$ axis & & & & & $b$ axis& & & & & $c$ axis & & & & \\
$\omega_{{\rm TO}j}$ & $\gamma_{{\rm TO}j}$ & $S_j$ &$\omega_{{\rm LO}j}$&$\gamma_{{\rm
LO}j}$& $\omega_{{\rm TO}j}$ & $\gamma_{{\rm TO}j}$ & $S_j$ &$\omega_{{\rm LO}j}$&$\gamma_{{\rm
LO}j}$ & $\omega_{{\rm TO}j}$  & $\gamma_{{\rm TO}j}$ & $S_j$ &$\omega_{{\rm
LO}j}$&$\gamma_{{\rm LO}j}$\\
\colrule
154 &  5.5  & 0.38 &155&4.3& 139 & 5.5  & 0.53 &141&4.0& 167 & 3.8 & 7.15 &196&2.9\\
209 &  5.1  & 3.77 &236&5.2& 240 & 8.8  & 2.36 &253&5.0& 204 & 4.9 & 0.30 &216&6.5\\
316 &  6.9  & 2.05 &331&3.6& 273 & 7.3  & 0.60 &279&9.8& 324 & 8.6 & 3.02 &346&4.8\\
335 &  3.1  & 0.12 &342&8.5& 308 & 5.3  & 0.51 &313&4.6& 349 & 6.2 & 0.08 &354&9.5\\
344 &  9.7  & 0.09 &351&6.6& 364\footnote{the splitting of this TO phonon
in $b$-axis is probably due to a cross talk of the $a$-axis with strong LO mode at 351 cm$^{-1}$.} & 10.0 & 1.37 &369&7.9& 387 & 9.0 & 2.37 &484&7.6\\
426 &  7.7  & 2.78 &508&13.2& 380 & 9.9  & 2.59 &438&6.9& 545 & 7.2 & 0.85 &695& 7.3\\
516 &  10.4 & 0.07 &525&7.5& 461 & 8.8  & 0.61 &496&9.2& 595 & 12.0& 0.01 &595 & 12.5\\
554 &  7.1  & 0.61 &683&7.1& 528 & 12.5 & 0.82 &689&10.6&  -- & --  & --   &--&--\\
576 &  11.3 & 0.02 &576&11.2& 578 & 11.2 & 0.02 &577&9.6&  -- & --  & --   &--&--\\
\colrule
$\epsilon_\infty$=3.5  & & &&&$\epsilon_\infty$=4.0
 & & &&&$\epsilon_\infty$=3.1 \\ $\epsilon^*_\infty$=4.0 & & &&&
$\epsilon^*_\infty$=4.1 & & &&& $\epsilon^*_\infty$=3.8
\end{tabular}
\end{ruledtabular}
\end{table*}

\begin{table*}
\caption
{Buckingham potential parameters for the shell-shell and core-shell interactions, the ionic shell charges (Y) (in units of a free electron charge) and force constants (k) in orthorhombic YTiO$_3$, r$_{cutoff}$ = 19 \AA.}
\begin{ruledtabular}
\begin{tabular}{llrllcc}   
   &  &  A(eV) &  &  $\rho $(\AA) &  & C(eV$\cdot $\AA $^{-6}$) \\
\colrule
\ O-O    & &22764.0  & &0.1490 & &20.37 \\
Ti-O1  & & 2902.8  & &0.2837 & &00.00 \\
Ti-O2  & & 1240.1  & &0.3313 & &00.00 \\
\ Y-O1 & & 3860.6  & &0.2944 & &00.00 \\
\ Y-O2 & & 12769.1  & &0.2495& &00.00 \\
\colrule
& Ion &  & Y ($\mid e \mid$) &  &k (eV $\cdot $\AA  $^{-2}$) \\
  & O1$^{2-}$  & &-2.000342    & &34.4 \\
  & O2$^{2-}$  & &-3.271589    & &35.7 \\
\end{tabular}
\end{ruledtabular}
\end{table*}

\begin{table*}
\caption{Experimental and calculated properties   of YTiO$_3$.
The lattice constants and fractional atomic coordinates present the data
for the Czochralski-grown single crystals at T = 293 K \cite{MacLean}.}
\begin{ruledtabular}
\begin{tabular}{ccc}   
Properties             &  Experiment & Calculation\\
\colrule
$a$, $\AA$   &  5.316(2) & 5.3124  \\
$b$, $\AA$   &  5.679(2) & 5.6880  \\
$c$, $\AA$   &  7.611(3) & 7.6072  \\
$V$, $\AA^3$ & 229.799   & 229.864 \\ \\

$x/a$, $y/b$, $z/c$  &           &     \\
\colrule \
 Y   &  0.9793(1),0.0729(1),0.25  & 0.9778,0.0692,0.25\\
 Ti  &  0.5,0.0,0.0               & 0.5,0.0,0.0     \\ 
 O1  &  0.121(1),0.458(1),0.25    & 0.1185,0.4485,0.25\\
 O2  &  0.691(1),0.310(1),0.058(1)& 0.6967,0.3055,0.0595\\ \\
 
Dielectric constants   &   & \\
\colrule \
$\epsilon^0_{11}$       & 13.7 & 13.34\\
$\epsilon^0_{22}$       & 13.6 & 14.02\\
$\epsilon^0_{33}$       & 17.7 & 17.52\\
$\epsilon^{\infty}_{11}$       & 3.5-4.0 & 3.84\\
$\epsilon^{\infty}_{22}$       & 4.0-4.1 & 4.31\\
$\epsilon^{\infty}_{33}$       & 3.1-3.8 & 3.21\\
\end{tabular}
\end{ruledtabular}
\end{table*}

\begin{table*}[tbp]
\caption{Calculated eigenfrequencies $\omega^{\rm SMC}_{{\rm TO}j}$ and components of eigenvectors $\chi_{ij}$ of the IR normal modes in orthorhombic YTiO$_3$. Signs of the components are referred to the atoms with fractional coordinates in the unit cell: Y(0.9793,0.0729,0.25), Ti(0.5,0,0), O1(0.121,0.458,0.25), and O2(0.691,0.310,0.058). The eigenvector  components used in the inversion of  Eq. (6) are underlined.}
\begin{ruledtabular}
\begin{tabular}{c ccc ccc ccc ccc}
  $\omega^{\rm SMC}_{{\rm TO}j}$ &\multicolumn{3}{c}{Y} & \multicolumn{3}{c}{Ti} & \multicolumn{3}{c}{O1} & \multicolumn{3}{c}{O2} \\
                          & $x$ & $y$ & $z$ & $x$ & $y$ & $z$ & $x$ & $y$ & $z$ &$x$ & $y$ & $z$ \\
\colrule
$B_{2u}$(1):149  & -0.378 & {0.042} & 0.000 & -0.265 & {-0.052} & -0.005 & 0.087 & {-0.005}  & 0.000 & 0.043 & {-0.001} & -0.104 \\
$B_{2u}$(2):203  &  0.015 & \uline{-0.221} & 0.000 &  -0.085 & \uline{0.373} & 0.094 & 0.005 & \uline{-0.052}  & 0.000 & 0.141 & \uline{-0.036} & 0.002  \\
$B_{2u}$(3):302  & 0.007 & {-0.178} & 0.000 & 0.043 &{0.057} & -0.157 & -0.157 & {0.030}  & 0.000 & -0.101 & {0.144} & -0.224  \\
$B_{2u}$(4):340  & 0.130 & \uline{0.059} & 0.000 & -0.099 & \uline{0.011} & 0.100 & 0.189 & \uline{-0.365}  & 0.000 & -0.083 & \uline{0.104} & -0.051  \\
$B_{2u}$(5):408  & 0.215 & {-0.084} & 0.000 & -0.143 & {-0.104} & 0.040 & 0.262 & {0.267}  & 0.000 & 0.063 & {0.055} & -0.071  \\
$B_{2u}$(6):446  & 0.117 & {0.115} & 0.000 & -0.041 & {0.011} & -0.381 & 0.035 & {-0.114}  & 0.000 & 0.142 & {-0.088} & -0.054  \\
$B_{2u}$(7):497  & -0.069 & {-0.148} & 0.000 & -0.092 & {0.011} & -0.222 & 0.072 & {0.002}  & 0.000 & -0.042 & {0.163} & 0.227  \\
$B_{2u}$(8):614  & -0.076 & \uline{-0.017} & 0.000 & 0.218 & \uline{-0.129} & 0.068 & 0.001 & \uline{-0.078}  & 0.000 & 0.235 & \uline{0.171} & -0.015  \\
$B_{2u}$(9):546  & 0.140 & {0.040} & 0.000 & -0.288 & {-0.109} & 0.076 & -0.327 & {-0.026}  & 0.000 & 0.072 & {0.061} & 0.038  \\
\colrule
$B_{3u}$(1):129  & {-0.049} & -0.310 & 0.000 & {0.114} & 0.308 & 0.044 & {0.005} & -0.073  & 0.000 & {-0.043} & 0.065 & 0.111    \\
$B_{3u}$(2):232  & \uline{-0.204} & 0.196 & 0.000 & \uline{0.305} & 0.065 & 0.177 & \uline{-0.059} & -0.013  & 0.000 & \uline{0.005} & 0.088 & -0.105    \\
$B_{3u}$(3): 291  & \uline{-0.223} & 0.109 & 0.000 & \uline{0.040} & -0.012 & -0.075 & \uline{0.119} & -0.055  & 0.000 & \uline{0.169} & -0.091 & 0.212   \\
$B_{3u}$(4):372  & {-0.141} & -0.116 & 0.000 &{0.047} & -0.113 & -0.239 & {0.289} & -0.061  & 0.000 & {-0.019} & 0.119 & -0.118   \\
$B_{3u}$(5):396  & {0.074} & 0.183 & 0.000 & {-0.158} & 0.174 & 0.099 &{0.188} & -0.324  & 0.000 & {-0.044} &-0.005 & -0.032   \\
$B_{3u}$(6):486  & {-0.089} & -0.147 & 0.000 & {-0.085} & 0.117 & 0.039 & {-0.038} & -0.019  & 0.000 & {0.197} &-0.139 & -0.200   \\
$B_{3u}$(7):495  & {-0.042} & -0.067 & 0.000 & {-0.100} & -0.059 & 0.330 & {0.267} & 0.220  & 0.000 & {0.002} &0.020 & 0.022   \\
$B_{3u}$(8):652  & \uline{0.028} & 0.100 & 0.000 & \uline{-0.180} & 0.142 & -0.067 & \uline{-0.062} & 0.119  & 0.000 & \uline{0.153} &0.242 & 0.013   \\
$B_{3u}$(9):561  & {0.004} & 0.143 & 0.000 & {0.029} & 0.266 & -0.171 & {0.123} & 0.265  & 0.000 & {-0.091} &-0.111 & -0.036\\
\colrule
$B_{1u}$(1):151  & 0.000 & 0.000 & \uline{-0.283} & -0.094 & 0.052 & \uline{0.167} & 0.000 & 0.000  & \uline{0.131} & 0.132 & -0.155 & \uline{0.123}    \\
$B_{1u}$(2):154  & 0.000 & 0.000 & {-0.012} & 0.064 & -0.442 & {-0.072} & 0.000 & 0.000  & {0.015} & -0.066 & -0.116 & {0.070}    \\
$B_{1u}$(3):322  & 0.000 & 0.000 & \uline{0.114} & 0.125 & 0.135 & \uline{-0.275} & 0.000 & 0.000  & \uline{-0.150} & 0.101 & -0.102 & \uline{0.179}   \\
$B_{1u}$(4):391  & 0.000 & 0.000 & {-0.153} & 0.385 & 0.092 & {0.102} & 0.000 & 0.000  & {0.030} & -0.148 & 0.040 & {0.077}   \\
$B_{1u}$(5):453  & 0.000 & 0.000 & {0.108} & 0.186 & 0.030 & {0.124} & 0.000 & 0.000  & {-0.065} & 0.067 & -0.214 & {-0.203}   \\
$B_{1u}$(6): 549  & 0.000 & 0.000 & \uline{-0.028} & -0.115 & -0.005 & \uline{0.202} & 0.000 & 0.000  & \uline{-0.415} & -0.082 & -0.024 & \uline{0.066}   \\
$B_{1u}$(7):599  & 0.000 & 0.000 & {-0.029} & 0.160 & -0.157 & {0.099} & 0.000 & 0.000  & {-0.108} & 0.244 & 0.170 & {0.003}   \\

\end{tabular}
\end{ruledtabular}
\end{table*}

\newpage
\begin{table*}
\caption{Suggested assignments of the TO phonon modes in the polarized ellipsometry
spectra, $\omega_{{\rm TO}j}$, with the normal modes resulting from the shell-model
lattice-dynamics calculations, $\omega^{\rm SMC}_{{\rm TO}j}$. Squares of net dipole moments calculated with the eigenvector components $\chi_{ij}$, the formal ionic charges $Q_i$ and the effective dynamical charges $Q^*_i$ are compared with the corresponding values estimated from the experimental phonon parameters
$\omega_{{\rm TO}j}$ and $S_j$.}
\begin{ruledtabular}
\begin{tabular}{cccccccc}
$\omega_{{\rm TO}j}(\omega^{\rm SMC}_{{\rm TO}j})$ & [$k\omega^2_{{\rm TO}j}S_j]$ & $[\sum_i Q_iu_{ij}]^2$ & $[\sum_i Q^*_iu_{ij}]^2$ &$\omega_{{\rm TO}j}(\omega^{\rm
SMC}_{{\rm TO}j})$ &
$[k\omega^2_{{\rm TO}j}S_j]$ & $[\sum_i Q_iu_{ij}]^2$ & $[\sum_i Q^*_iu_{ij}]^2$\\
\colrule
$B_{3u}$(1):154\ (129)  &  0.04   & 0.09 & 0.20\(\pm0.02\) &$B_{2u}$(1):139\ (149) & 0.05  & 0.001 & 0.001\\
$B_{3u}$(2):209\ (232)  &  0.78   & 0.13 & 0.78\(\pm0.06\) &$B_{2u}$(2):240\ (203) & 0.64   & 0.38& 0.64\(\pm0.05\)\\
$B_{3u}$(3):316\ (291)  &  0.97   & 1.27 & 0.97\(\pm0.27\) &$B_{2u}$(3):273\ (302) & 0.21   & 0.59& 1.04\(\pm0.20\)\\
$B_{3u}$(4):335\ (372)  &  0.06   & 0.36 & 0.40\(\pm0.17\) &$B_{2u}$(4):308\ (340) & 0.23 & 0.17 & 0.23\(\pm0.04\)\\
$B_{3u}$(5):344\ (396)  &  0.05   & 0.14 & 0.53\(\pm0.10\) &$B_{2u}$(5):364\ (408) & 0.86 &1.08  & 1.78\(\pm0.30   \) \\
$B_{3u}$(6):426\ (486)  &  2.38  & 0.95  & 0.91\(\pm0.17\) &$B_{2u}$(6):380\ (446) & 1.76   & 0.56 & 0.95\(\pm0.16\)  \\
$B_{3u}$(7):516\ (495)  & \ \ 0.09    & 0.59 & 1.03\(\pm0.27\) &$B_{2u}$(7):461\ (497) & 0.61   & 0.68& 1.22\(\pm0.20\)\\
$B_{3u}$(8):554\ (652)  & \ 0.88   & 0.59 & 0.88\(\pm0.12\) &$B_{2u}$(8):528\ (614) & 1.08   & 0.60& 1.08\(\pm0.17\)\\
$B_{3u}$(9):576\ (561)  & \ \ 0.03    & 0.03 & 0.01\(\pm0.003\) &$B_{2u}$(9):578\ (546) & 0.03   & 0.11& 0.20\(\pm0.03\)\\ \colrule
$B_{1u}$(1):167\ (151)  &  0.94  & 0.68 & 0.94\(\pm0.15\)\\
$B_{1u}$(2):204\ (154)  &  0.06  & 0.20 & 0.32\(\pm0.05\)\\
$B_{1u}$(3):324\ (322)  &  1.50  & 0.56 & 1.50\(\pm0.20\)\\
$B_{1u}$(4):349\ (391)  &  0.05  & 0.15 & 0.21\(\pm0.03\)\\
$B_{1u}$(5):387\ (453)  &  1.67  & 1.68 & 3.19\(\pm0.50\)\\
$B_{1u}$(6):545\ (541)  &  1.19  & 0.78 & 1.19\(\pm0.20\)\\
$B_{1u}$(7):595\ (599)  &  0.02  & 0.12 & 0.21\(\pm0.03\)\\
\end{tabular}
\end{ruledtabular}
\end{table*}

\begin{table*}
\caption{The dynamical effective charges of the atoms (in units
of a free electron charge) calculated from experimental phonon parameters,
$\omega_{{\rm TO}j}$ and $S_j$, using the  eigenvector components $\chi_{ij}$ listed in Table V. Uncertainties of the effective charges are estimated by varying the experimental oscillator strength by \(\pm10\% \).}
\begin{ruledtabular}
\begin{tabular}{ccccc}
    & $Q^*_{Y}$  & $Q^*_{Ti}$ & $Q^*_{O1}$ & $Q^*_{O2}$\\
\colrule
$a$-axis: $B_{3u}$(2)\{209\}+ $B_{3u}$(8)\{554\}\ &   & &  & \\
+ $B_{3u}$(3)\{316\}\    &  1.5\(\pm0.3\) & 5.0\(\pm0.2\) & -2.6\(\pm0.4\) & -1.9\(\pm0.2\) \\
\colrule
$b$-axis: $B_{2u}$(2)\{240\}+ $B_{2u}$(8)\{528\}\ &   &  &   &  \\
+ $B_{2u}$(4)\{308\}\    &  3.9\(\pm0.4\)\(\)  & 3.9\(\pm0.2\)\(\) & -2.5\(\pm0.2\)\(\) & -2.7\(\pm0.2\)\(\) \\
\colrule
$c$-axis: $B_{1u}$(1)\{167\}+ $B_{1u}$(6)\{545\}\  &  &  &  & \\
 + $B_{1u}$(3)\{324\}\    &  3.4\(\pm0.2\) & 4.6\(\pm0.3\) & -2.3\(\pm0.2\) & -2.8\(\pm0.2\) \\
\end{tabular}
\end{ruledtabular}
\end{table*}

\newpage

\begin{figure}[tbp]
\includegraphics*[width=170mm]{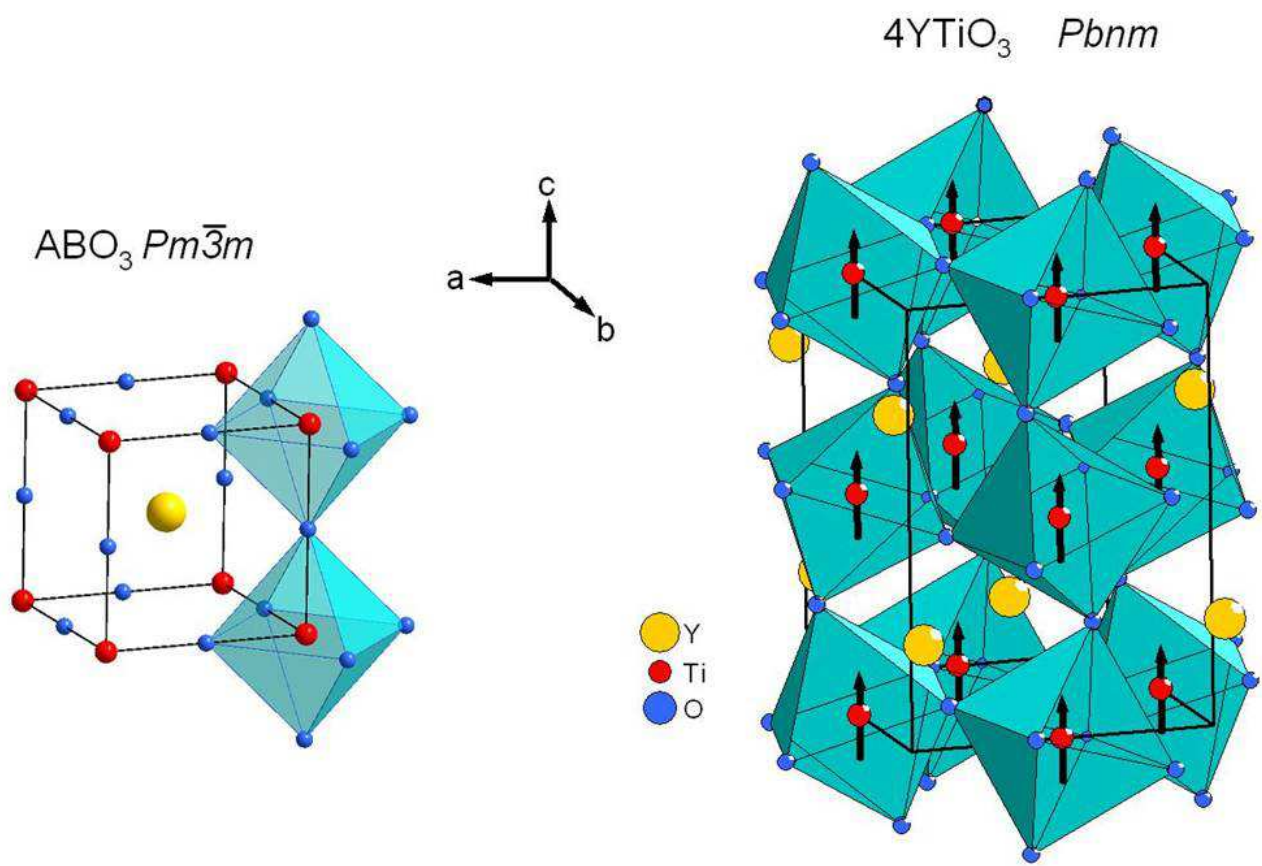}
\caption{Crystal structure of cubic perovskite oxides ABO$_3$ (space group $Pm\bar3m$) where A is a rare-earth atom (yellow sphere) and
B is a transition-metal atom (red spheres) (left), and related orthorhombic
structure of YTiO$_3$ (space group $Pbnm$), with ferromagnetic alignment
of Ti spins shown schematically (right).}
\label{Fig1}
\end{figure}

\begin{figure}[tbp]
\includegraphics*[width=120mm]{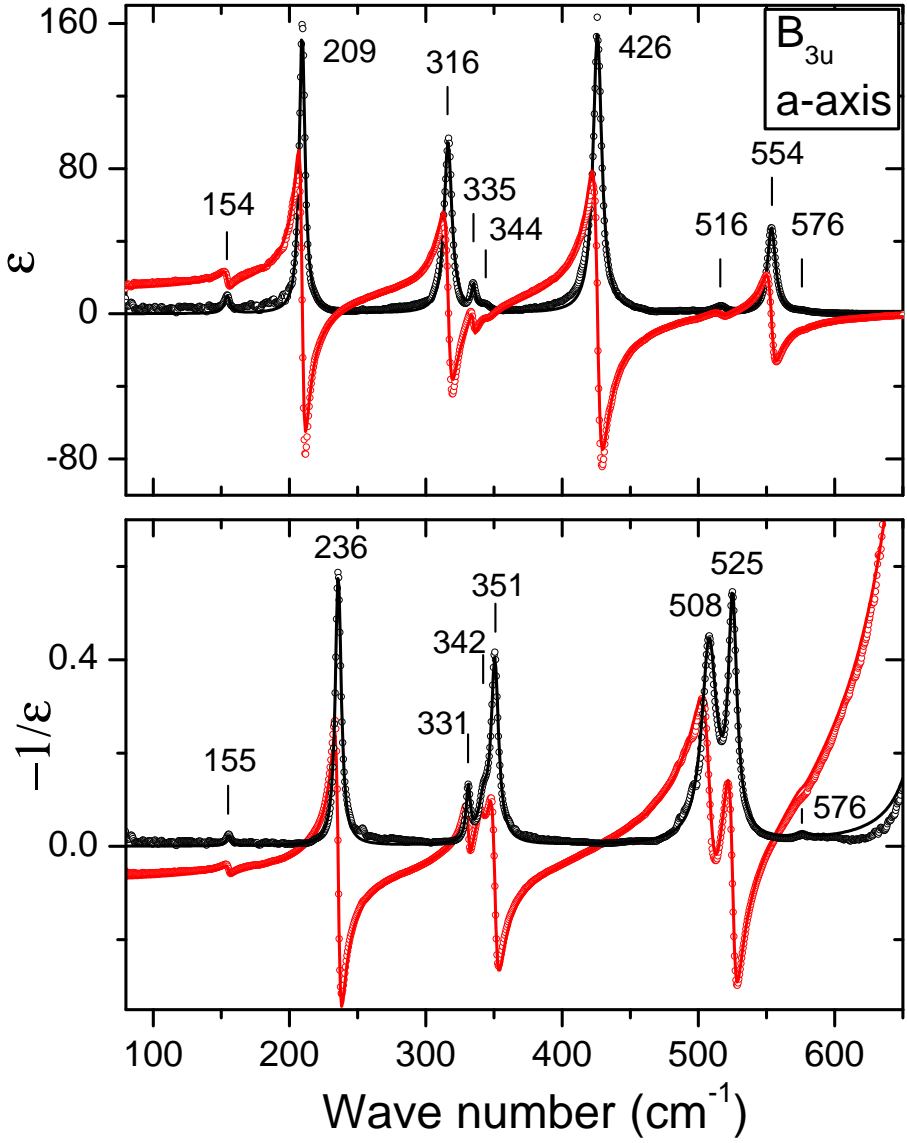}
\caption{Real (red open circles) and imaginary (black open circles) parts
for the measured $a$-axis complex dielectric response $\tilde\epsilon(\omega)$ and its inverse -1/$\tilde\epsilon(\omega)$. The solid curves present the result of the fitting with a set of Lorenzian oscillators, which resonant frequencies $\omega_{{\rm TO}j}$ and $\omega_{{\rm LO}j}$ are indicated (see also Table II).}
\label{Fig2}
\end{figure}

\begin{figure}[tbp]
\includegraphics*[width=120mm]{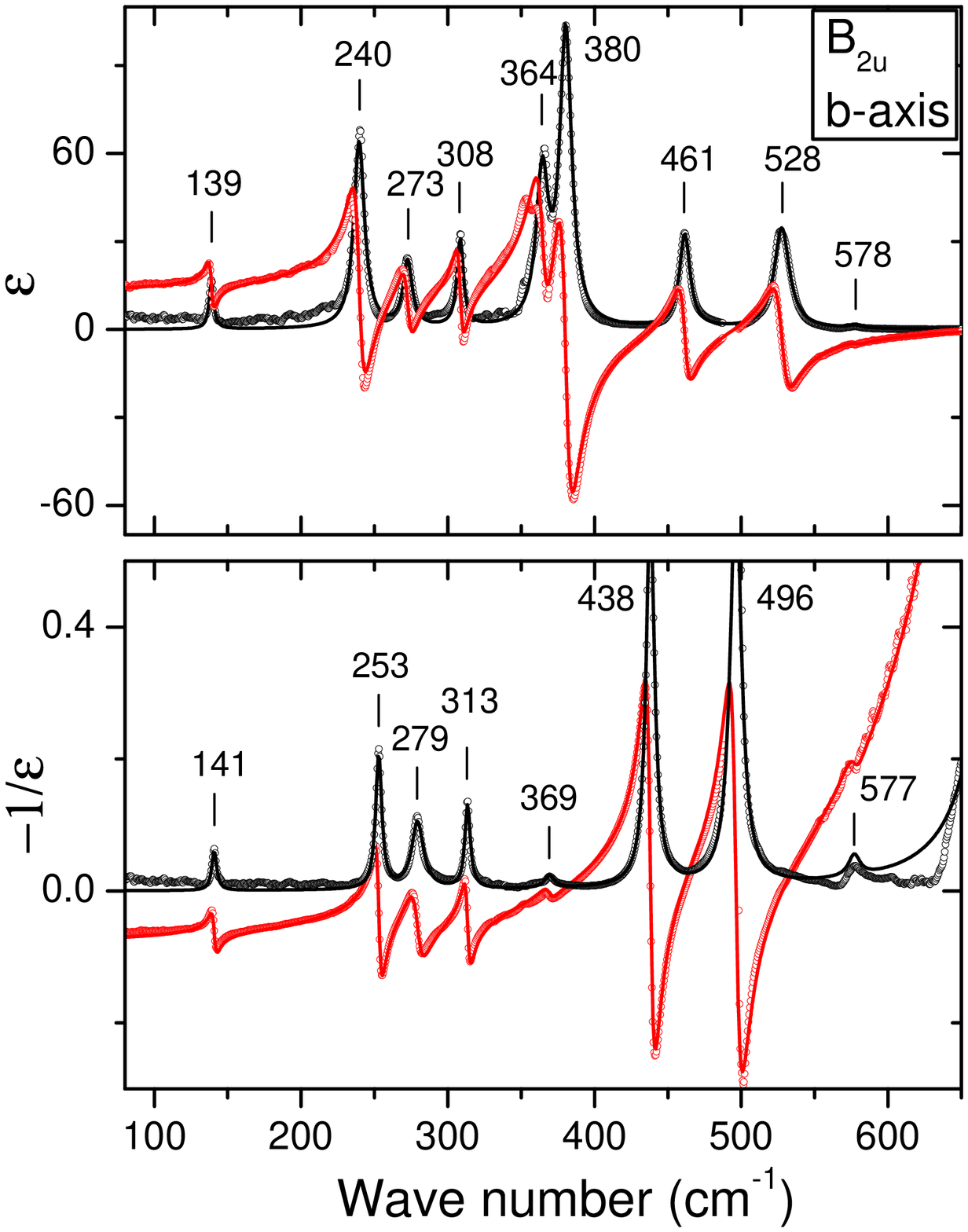}
\caption{The same as in Fig. 2 for the $b$-axis complex dielectric response $\tilde\epsilon(\omega)$ and its inverse -1/$\tilde\epsilon(\omega)$.}
\label{Fig3}
\end{figure}

\begin{figure}[tbp]
\includegraphics*[width=120mm]{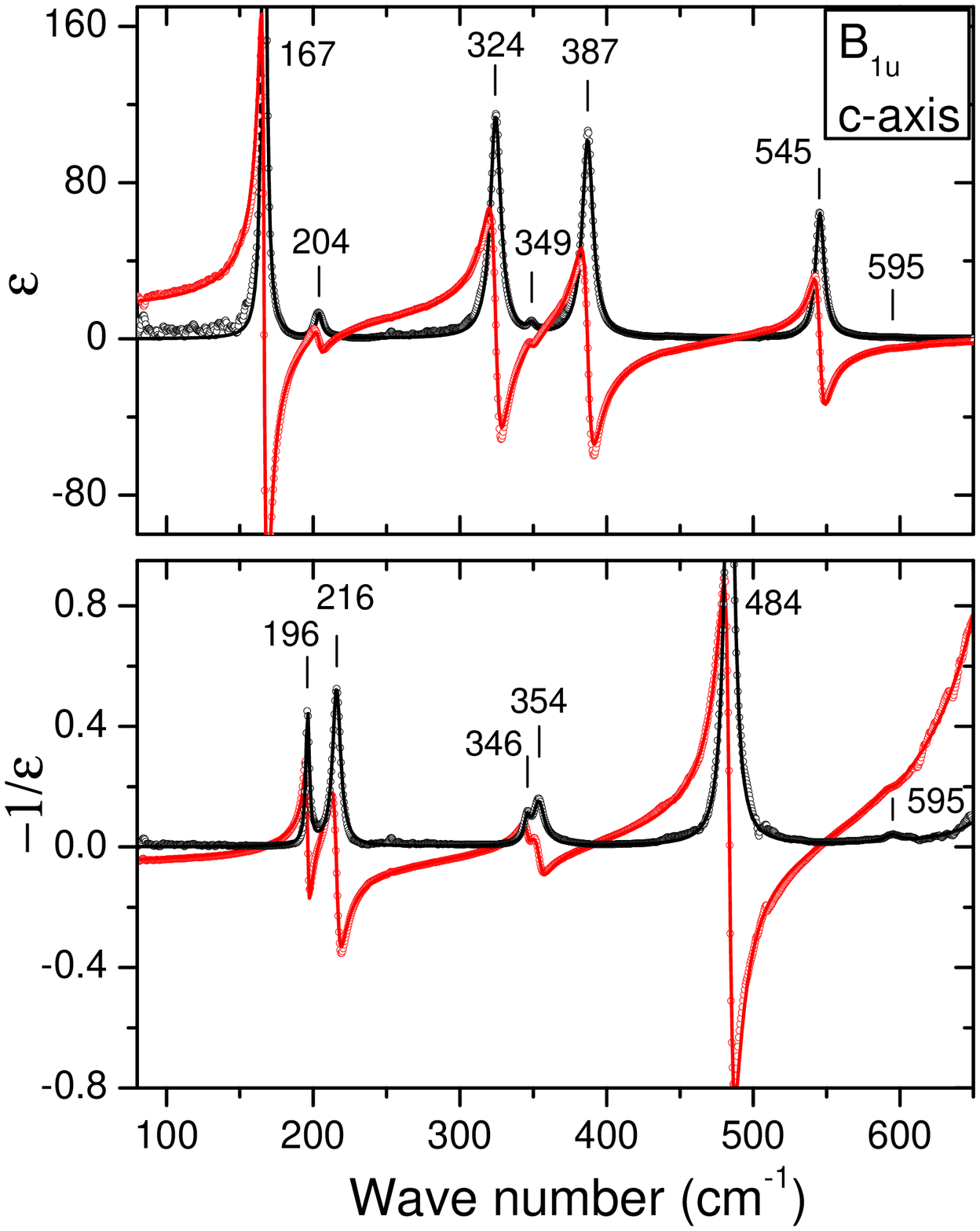}
\caption{The same as in Fig. 2 for the $c$-axis complex dielectric response $\tilde\epsilon(\omega)$ and its inverse -1/$\tilde\epsilon(\omega)$.}
\label{Fig4}
\end{figure}

\begin{figure}[tbp]
\includegraphics*[width=150mm]{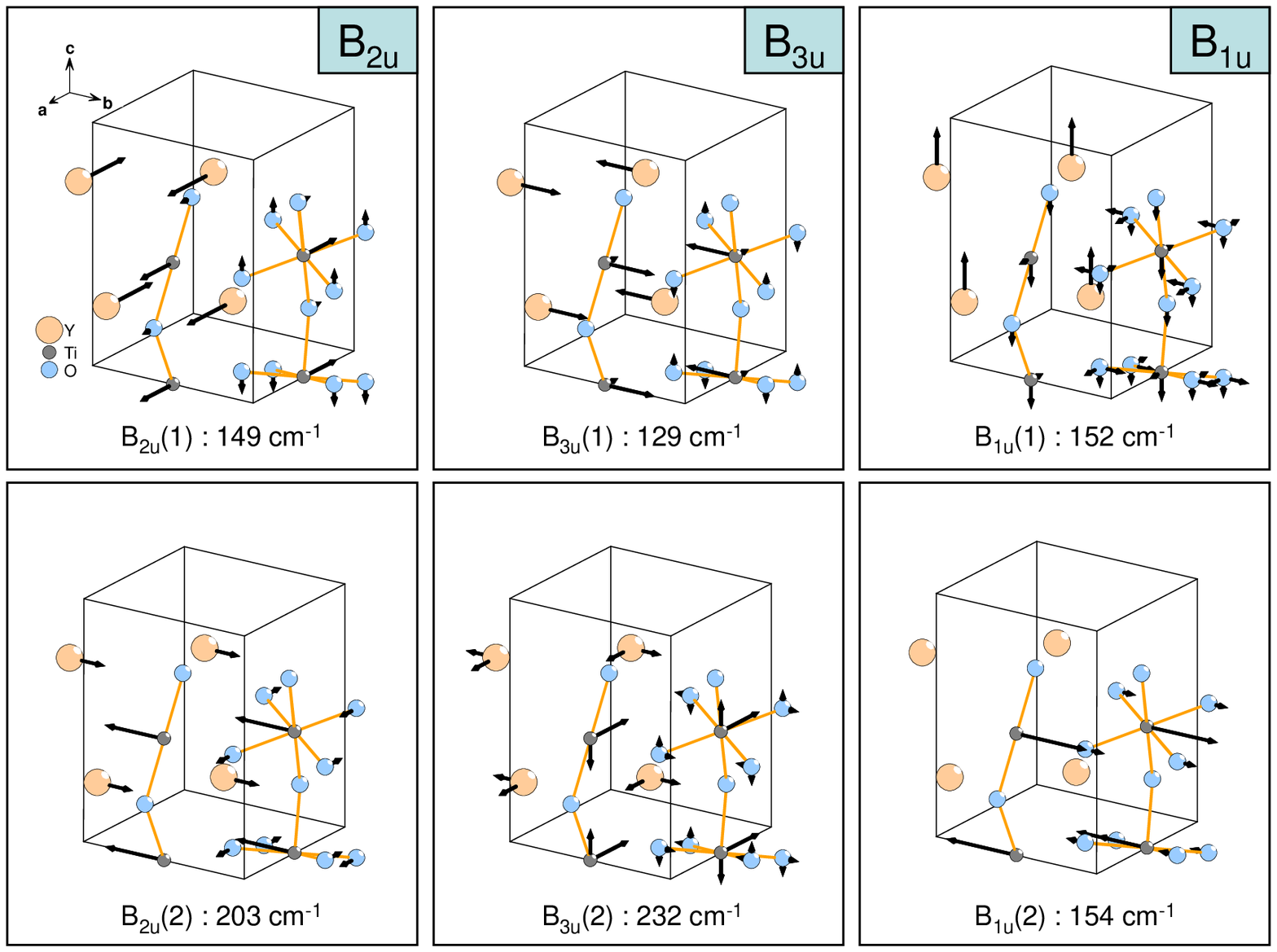}
\caption{Eigenfrequencies and eigenvector components of the normal modes of $B_{1u}$, $B_{2u}$, and $B_{3u}$ symmetries in orthorhombic YTiO$_3$ at low frequencies, near the characteristic frequency of the perovskite $external$ mode.}
\label{Fig5}
\end{figure}

\begin{figure}[tbp]
\includegraphics*[width=150mm]{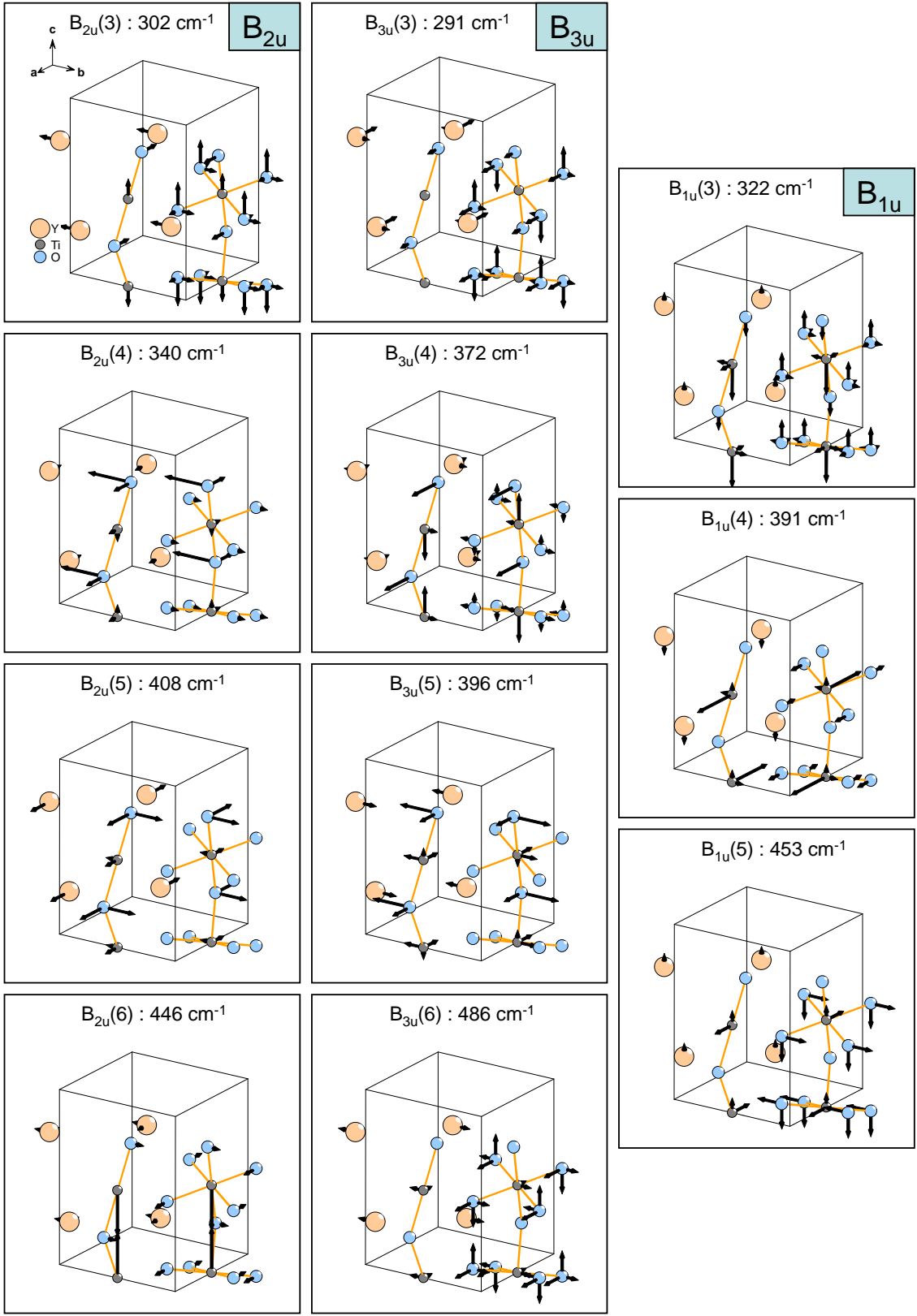}
\caption{Eigenfrequencies and eigenvector components of the normal modes of $B_{1u}$, $B_{2u}$, and $B_{3u}$ symmetries in orthorhombic YTiO$_3$ near the characteristic frequency of the perovskite $bending$ mode.}
\label{Fig6}
\end{figure}

\begin{figure}
\includegraphics*[width=150mm]{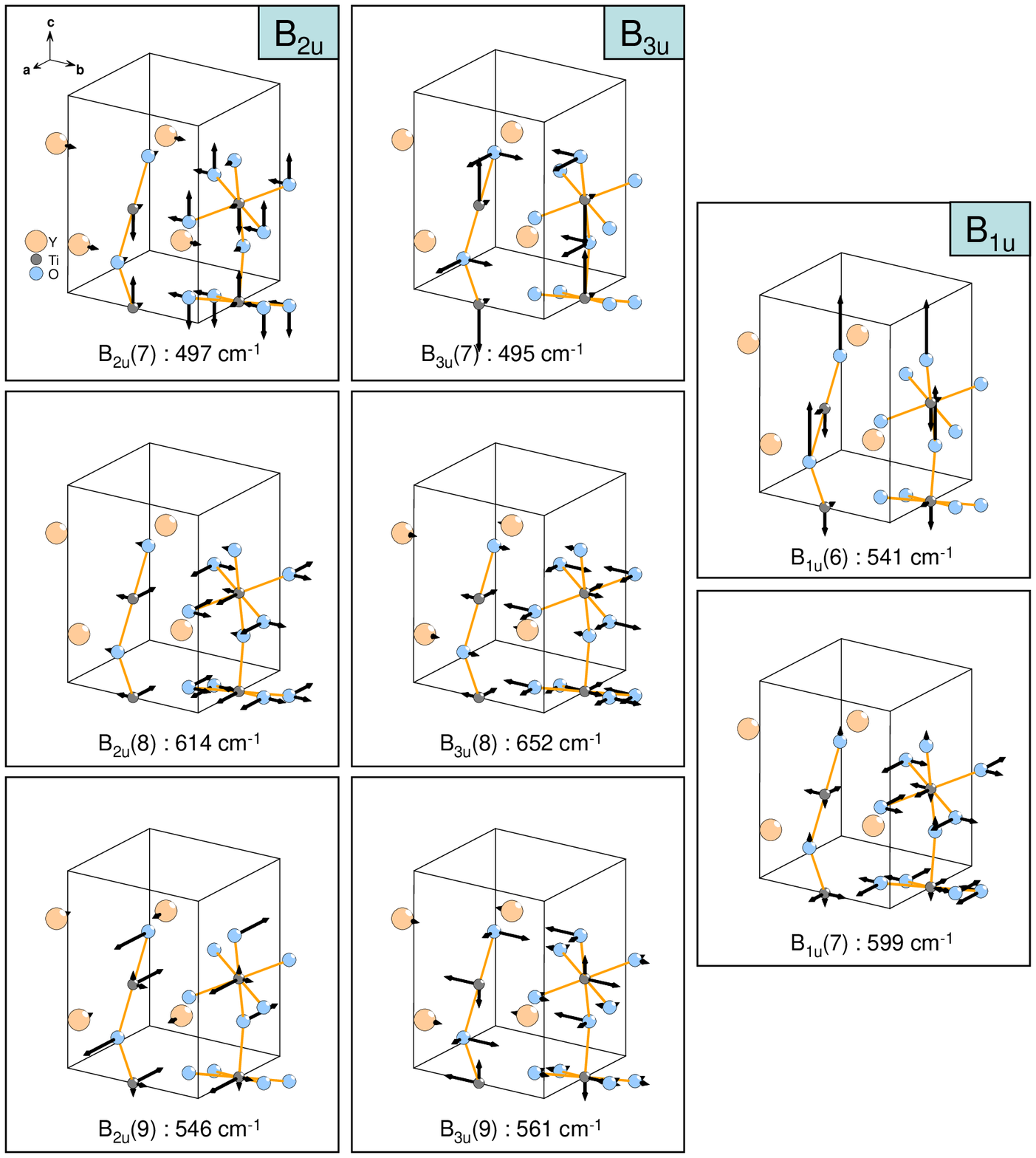}
\caption{Eigenfrequencies and eigenvector components of the normal modes of $B_{1u}$, $B_{2u}$, and $B_{3u}$ symmetries in orthorhombic YTiO$_3$ at high frequencies, near the characteristic frequency of the perovskite $stretching$ mode.}
\label{Fig7}
\end{figure}


\begin{thebibliography}{99}
\bibitem{note} To extract information about the phonons we fitted the ellipsometry spectra in Figs. 2-4 using a simple Lorentz model for noninteracting harmonic oscillators. However, as one can see from Figs. 2-4, several phonon bands show weak asymmetry, with flatter rise and steeper fall of the imaginary
part of $\tilde\epsilon(\omega)$. At the same time the $-1/\tilde\epsilon(\omega)$ spectra were found to be fitted fairly well with a Lorentz model. The asymmetry of optical phonon modes suggests mode coupling and interaction damping, which can be accounted for by a phenomenological model for coupled oscillators, $\tilde\epsilon(\omega) = \epsilon_{\infty}+\sum_j S_j\frac{\omega^2_j-i\omega \sigma_j}{\omega_j^2-\omega^2-i\omega\gamma_j}$, with the asymmetry parameters
$\sigma_j$ \cite{Humlicek}. We estimate that the level of asymmetry of all TO phonons in YTiO$_3$ is fairly weak and within the experimental uncertainties
in our case.
\end{thebibliography}
\end{document}